\documentclass[english,twocolumn,superscriptaddress,
  longbibliography]{revtex4-1}

\usepackage{soul}
\usepackage[dvipsnames]{xcolor}

\usepackage{xcolor}
\usepackage{graphicx}
\usepackage{amsmath}
\usepackage{amsfonts}
\usepackage{color}
\usepackage{babel}
\usepackage{url}
\usepackage{hyperref}
\usepackage{tikz}
\usepackage{CJKutf8}
\usepackage[title]{appendix}
\usepackage{siunitx}

\definecolor{lime}{HTML}{A6CE39}
\DeclareRobustCommand{\orcidicon}{
	\begin{tikzpicture}
	\draw[lime, fill=lime] (0,0) 
	circle [radius=0.16] 
	node[white] {{\fontfamily{qag}\selectfont \tiny ID}};
	\draw[white, fill=white] (-0.0625,0.095) 
	circle [radius=0.007];
	\end{tikzpicture}
	\hspace{-2mm}
}

\foreach \x in {A, ..., Z}{\expandafter\xdef\csname
  orcid\x\endcsname{\noexpand\href{https://orcid.org/\csname
      orcidauthor\x\endcsname} {\noexpand\orcidicon}} }

\begin{document}

\title{Online regularization of Poincar\'e map of storage rings with Shannon entropy}

\author{Yongjun Li\orcidA{}}\thanks{email: yli@bnl.gov}
\affiliation{Brookhaven National Laboratory, Upton, New York 11973, USA}
\author{Kelly Anderson}\affiliation{Michigan State University, East Lansing, Michigan  48864, USA}
\author{Derong Xu\orcidC{}}\affiliation{Brookhaven National Laboratory, Upton, New York 11973, USA}
\author{Yue Hao}\affiliation{Michigan State University, East Lansing, Michigan  48864, USA}
\author{Kiman Ha}\affiliation{Brookhaven National Laboratory, Upton, New York 11973, USA}
\author{Yoshiteru Hidaka}\affiliation{Brookhaven National Laboratory, Upton, New York 11973, USA}
\author{Minghao Song}\affiliation{Brookhaven National Laboratory, Upton, New York 11973, USA}
\author{Robert Rainer\orcidB{}}\affiliation{Brookhaven National Laboratory, Upton, New York 11973, USA}
\author{Victor Smaluk}\affiliation{Brookhaven National Laboratory, Upton, New York 11973, USA}
\author{Timur Shaftan}\affiliation{Brookhaven National Laboratory, Upton, New York 11973, USA}
 
\begin{abstract}
    Shannon entropy, as a chaos indicator, is used for online Poincar\'e map regularization and dynamic aperture optimization in the National Synchrotron Light Source-II (NSLS-II) ring. Although various chaos indicators are widely used in studying nonlinear dynamical systems, including modern particle accelerators, it is the first time to use a measurable one in a real-world machine for online nonlinear optimization. Poincar\'e maps, constructed with the turn-by-turn beam trajectory readings from beam position monitors, are commonly used to observe the nonlinearity in ring-based accelerators. However, such observations typically only provide a qualitative interpretation. We analyze their entropy to quantify the chaos in measured Poincar\'e maps. After some canonical transformations on the Poincar\'e maps, not only can the commonly used nonlinear characterizations be extracted, but more importantly, the chaos can be quantitatively calibrated with Shannon entropy, and then used as the online optimization objectives.
\end{abstract}
 
\maketitle

 Modern ring-based accelerators need strong nonlinear magnets to mitigate the chromatic aberrations and maintain beam stability. However, the nonlinearity causes the dynamics to depend on the amplitude of particles and experience nonlinear resonances. As a result, particles can only maintain long-term stability within a region which is referred to as dynamic aperture (DA). It is generally believed that a nonlinear lattice can be optimized through mitigating the chaotic motion of particles. Various chaos indicators have been explored over time~\cite{bazzani2023} such as: frequency map analysis~\cite{laskar2003frequency,papaphilippou2014detecting}, the maximal Lyapunov exponent~\cite{todesco1996}, nonlinear resonance driving terms extracted from Hamiltonian perturbation theory~\cite{guignard1985,cai2020,dragt2020}, time-reversal errors~\cite{li2021} etc. These are widely used in design simulation and offline optimization. However, these methods are difficult to use for online machine tuning.
 
 We report a method of chaos characterization from measured recurrent Poincar\'e maps~\cite{poincare1967new} in storage rings. The Poincar\'e maps are formed from the intersections of multi-turn phase trajectories within a designated transverse cross-section. Experimentally, they can be constructed with beam position monitors (BPM) with a turn-by-turn resolution after the beam is being excited. So far, Poincar\'e maps mainly provide a visual observation rather than a quantitative measure of chaos. We introduce a canonical coordinate, in which chaos can be conveniently analyzed with Shannon entropy~\cite{shannon1948}. With this analysis, not only can the commonly used tune-shift-with-amplitude (TSWA) and action-angle smears be extracted, but more importantly, the overall chaos of nonlinear beam motion can be characterized. As a proof of concept, we experimentally demonstrate how Poincar\'e maps can be regularized with the entropy as a chaos indicator in the NSLS-II storage ring.

\section*{Methods}
 In information theory, Shannon entropy~\cite{shannon1948} is used to quantify the uncertainty or unpredictability in a random variable. Due to the common unpredictable behavior in nonlinear dynamical systems, entropy can be applied to study the chaos~\cite{letellier2006}. In a Poincar\'e map formed with a particle's $N$-turn coordinate $\mathbf{X}=(x,p_x,y,p_y,z,p_z)^T$, the entropy is defined as
 \begin{equation}\label{eq:shen1}
     S = -\int_{-\infty}^{+\infty}p(\mathbf{X})\log_2 p(\mathbf{X}) \text{d}\mathbf{X}.
 \end{equation}
 Since a linear regular motion on an invariant torus is completely predicable, its entropy should be zero naturally. But if computing with Eq.~\eqref{eq:shen1} in the conventional position-momentum coordinate, the entropy unnecessarily depends on $N$. To address this limitation, we propose applying the following canonical transformations prior to estimating entropy. Assuming the initial condition at a specific longitudinal location is $\mathbf{X}_0$. After a full turn it becomes $\mathbf{X}_1=\mathbf{MX}_0$, with the one-turn transportation matrix $\mathbf{M}$. If the motion is stable, a set of Twiss parameters can be defined to normalize the coordinates into a set of conjugate action-angle variables $(J_x,\phi_x,J_y,\phi_y,J_z,\phi_z)$~\cite{courant1958}. In this action-angle coordinate, the linear motion is purely rotational. It has fixed radii $J_{x,y,z}$ and phase advances $\Delta\phi_{x,y,z}=2\pi\nu_{x,y,z}$ in each plane. $\nu_{x,y,z}$ are known as the linear tunes. Now we introduce a parallel-shift transformation to each angle variable, the new variables read as  
 \begin{equation}\label{eq:angle}
     \hat{J}=J,\;\hat{\phi}_n =\phi_n-n\times 2\pi\overline{\nu},
 \end{equation}
 where $n=0,1,2,\cdots$ is the index of turns, and $2\pi\overline{\nu}$ is the average phase advance per turn. By applying the above transformations, a linear motion is mapped to a single state, with zero entropy correspondingly. If a linear coupling exists there, a decoupling process through similar transformations such as Edward-Teng's parameterization~\cite{edwards1973}, would be needed first. In principle, the transverse Poincar\'e maps in accelerators are 4D tori. Since the coupling is weak, or can be decoupled in our study, we can use their 2D projections in the horizontal and vertical planes respectively.

\begin{figure}
     \centering
     \includegraphics[width=1\columnwidth]{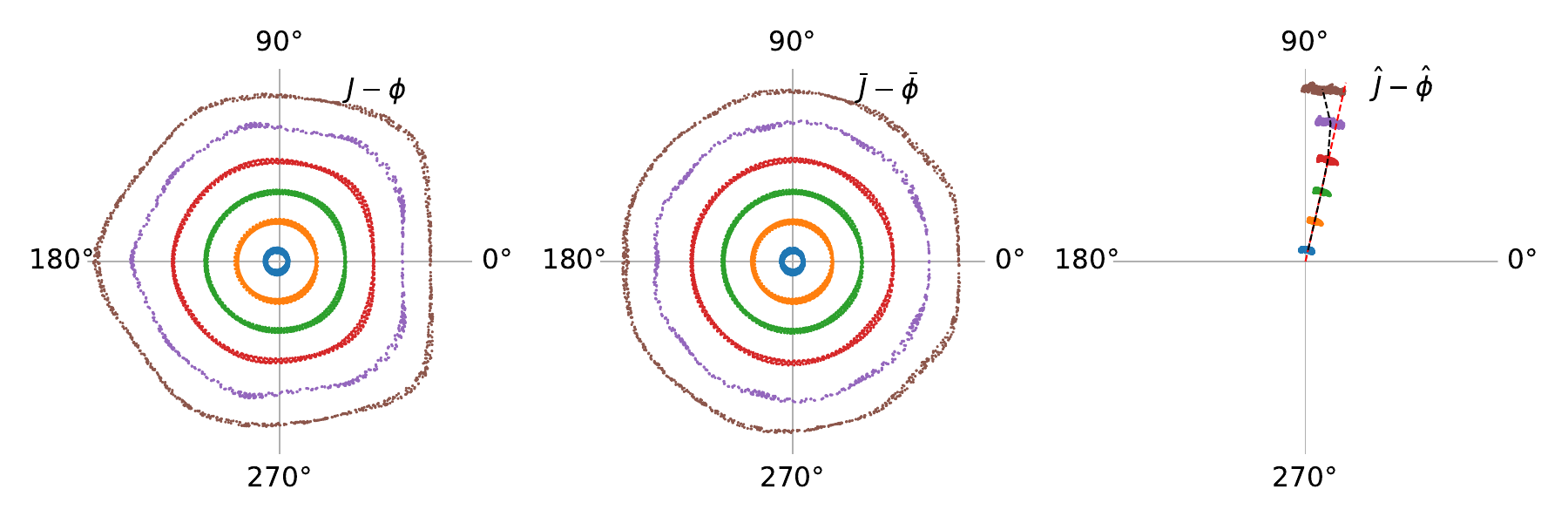}
     \caption{Poincar\'e maps through a sequence of canonical transformations for chaos characterization. Left: maps with the Courant-Synder action-angle. Middle: maps after a normal form transformation with Eq.~\eqref{eq:normal_form}. Right: maps after a parallel-shift of angle variable with Eq.~\eqref{eq:angle}. Due to the TSWA effect, the island centroids (black dashed curve) gradually deviate from the linear tune (red dashed ray).}
     \label{fig:canonical_trans}
 \end{figure}

 When nonlinear magnets are added to a ring lattice, the particle's coordinate typically does not precisely overlap its previous state due to nonlinear effects such as TSWA and action-angle smearing. The TSWA $(\frac{\partial \nu}{\partial J})$ describes an average phase-advance shift with action amplitude. The distortions of action-angles are quantified by the smear parameters~\cite{chao1988}, i.e., their fractional deviations $(\frac{\Delta J}{J},\,\frac{\Delta\phi}{\phi})$. From a given initial condition, a new state is evolved after each turn, with the exception of some special points such as the origin and fixed points. Its Poincar\'e map gradually diffuses with the number of passes. According to the Kolmogorov-Arnold-Moser theorem, when the fractional tune is irrational, the quasi-periodic motion of the particle is distorted but not destroyed. In such instances, a normal form technique~\cite{bazzani1994normal} is used to transfer the map into a particular simple form. Normal form is a generalized Courant-Synder normalization~\cite{courant1958} through a nonlinear canonical transformation. The new map usually has explicit invariants and more symmetries. Therefore, for nonlinear motions, prior to the parallel-shift transformation of Eq.~\eqref{eq:angle}, we also implement another normal form with a type-2 generating function
 \begin{equation}\label{eq:generating_fun}
     G_2(\phi,\bar{J})= \phi\bar{J}+\sum_{m=1}^kC_m\sin(m\phi+\phi_{m,0}),
 \end{equation}
 and the new action-angle variables become
 \begin{equation}\label{eq:normal_form}
    \bar{\phi}=\phi,\;
    \bar{J}=J+\sum_{m=1}^kmC_m\cos(m\phi+\phi_{m,0}). 
 \end{equation}
 The coefficients $C_m,\,\phi_{m,0}$ are chosen to minimize the variance of $\bar{J}$. With this normal form, only the fluctuations around the new $\bar{J}$ are taken into account when estimating its chaos. After a sequence of canonical transformations as illustrated in Fig.~\ref{fig:canonical_trans}, the final poincar\'e maps shrink into islands. Its dimensions in the radial and azimuth directions represent its action-angle smears, respectively. Obviously, the angular deviation of the island centroids (black dashed ray) from the linear motion (red dashed ray) represents the TSWA effect.

  With the above transformations, we can estimate the entropy of the island-shape Poincar\'e maps in the $\hat{J}-\hat{\phi}$ plane (illustrated in the right subplot in Fig.~\ref{fig:canonical_trans}). Usually the length of a valid multi-turn data acquired by BPMs is limited to a few hundreds turns due to the decoherence effect~\cite{meller1987decoherence,lee1991decoherence} and radiation damping (for lepton machines). In this case, measured distribution function $p(\hat{J},\hat{\phi})$ is discontinuous. It is practical to use a discrete distribution rather than Eq.~\eqref{eq:shen1} to estimate the entropy. The $\hat{J}-\hat{\phi}$ plane is divided into a mesh grid with each grid representing a macro-state, as denoted with an index ``$i$". When $N$-turn maps are projected onto the mesh grid, and more than two points are located in the same grid, they are considered to have the same macro-state~\cite{hao1991symbolic}. Each macro-state's probability is counted with $p_i=\frac{N_i}{N}$. Here $N_i$ is the number of the $i^{th}$ macro-state. The entropy $S$ can be estimated by
  \begin{equation}\label{eq:shen2}
      S=-\sum_i{p_i\log_2(p_i)}.
  \end{equation}
  When a nonlinear motion's action-angle diffuses out of its initial grid, the number of macro-states is greater than 1, which leads to an increment in the entropy. Given a reasonable grid dimension, the entropy quantitatively describes the chaos of a particle's motion. The dimension of the mesh grids is critical for entropy estimation. The grid should be sufficiently small to resolve the evolution of macro-states, and also large enough to compromise to our measurement precision. 
  Based on the dimension of measured islands and the BPM turn-by-turn resolution, the dimension of grid is chosen as $\Delta\hat{J}\times\Delta\hat{\phi}=1\times10^{-7}\,m\times1^{\circ}$, which gives the smallest error-bars in repeated measurements.
  
\begin{figure}[!ht]
    \centering
    \includegraphics[width=\columnwidth]{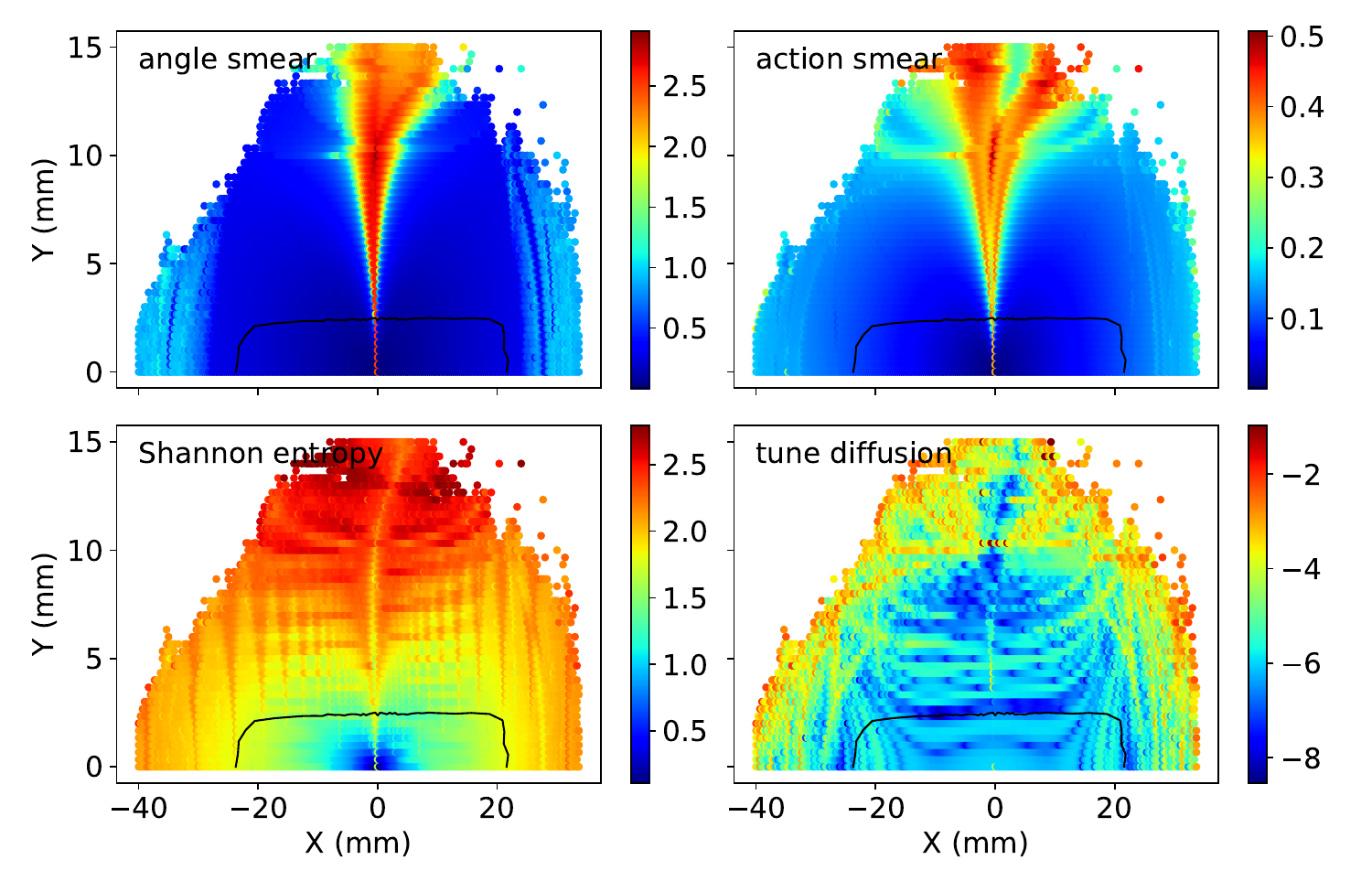}
    \caption{Comparison of action-angle smears, entropy and frequency map analysis for the NSLS-II ring. The black solid lines are the realistic DA when multipole errors and the physical aperture are taken into account.}
    \label{fig:smear_shen_fma}
  \end{figure}

  Action-angle smears are also proportional to the degree of chaos, but we observe that they can provide much less information than entropy. A comparison among smears, entropy and frequency map analysis (FMA) is made for the NSLS-II lattice, as shown in Fig.~\ref{fig:smear_shen_fma}.  The chaos pattern seen by FMA is similar to the Shannon Entropy. From the point of view of simulation, entropy might not have an obvious advantage. From a more practical standpoint, however, entropy can be obtained with less data and doesn't depend on the precise frequency evaluation as needed by FMA, which is beneficial for experimental measurements in electron machines. Due to synchrotron radiation and beam decoherence, a near constant amplitude is difficult to maintain for high resolution tune diffusion ($\Delta\nu\leq10^{-8}$) measurement, particularly for large amplitude oscillations. Another benefit is that the entropy varies monotonically with amplitude inside the DA, while tune diffusion has a fluctuation when crossing weak resonances. Therefore, using the tune diffusion as a metric could result in displacing weak resonances, rather than mitigating the chaos.
  
\section*{Results}
  The entropy-based online nonlinear phase space optimization has been demonstrated on the NSLS-II storage ring, a dedicated $3^{rd}$ generation medium energy light source~\cite{dierker2007}. Its main storage ring’s lattice has the traditional double-bend-achromat (DBA) structure. Due to various imperfections and hardware constraints, the measured DA is somewhat smaller than the design model. Sometimes, the off-axis injection efficiency is slightly lower than our expectation. We would like to use six families of harmonic sextupoles, located at dispersion-free sections, as the tuning knobs to reduce the chaos and improve the injection efficiency. Under different sextupole configurations, a stored beam is excited to four different amplitudes with a fast pulse magnet (so-called ``pinger") in the transverse plane. Poincar\'e maps are constructed with the turn-by-turn data seen by a pair of BPMs, which are recently installed in one of the magnet-free straight sections. Such BPM configuration grants us precise measurement of the Poincar\'e maps there~\cite{li2024}. The Twiss parameters used for the Courant-Synder normalization are also measured, but at sufficiently small excitation amplitudes. In this experiment, no accelerator model is needed as the entropy measurement is completely model-independent.
  
\begin{figure}[!ht]
     \centering
     \includegraphics[width=0.6\columnwidth]{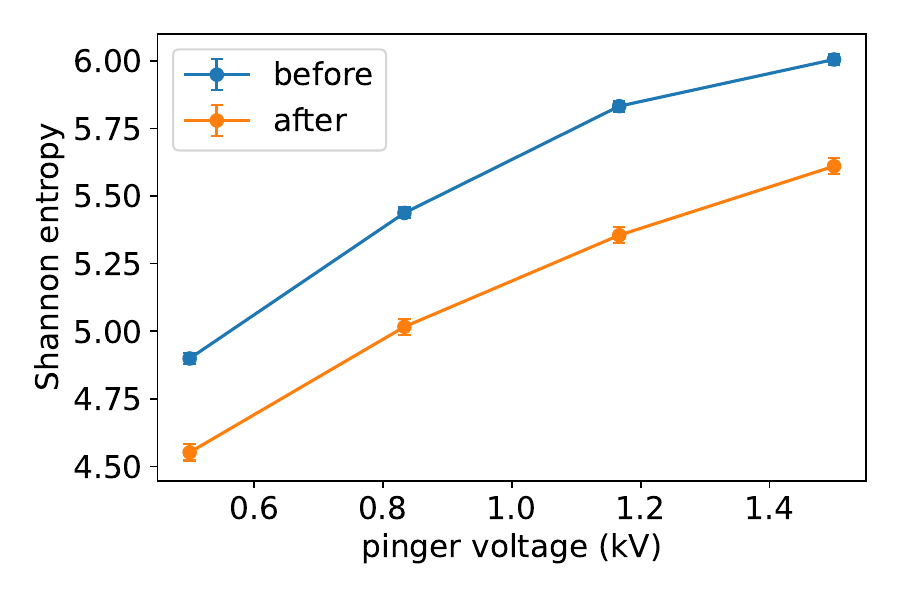}
     \caption{Entropy reduction at different oscillation amplitudes (corresponding to pinger excitation voltages) after an online sextupole optimization.}
     \label{fig:ShenOpt}
 \end{figure}

\begin{figure}[!ht]
     \centering
     \includegraphics[width=0.8\columnwidth]{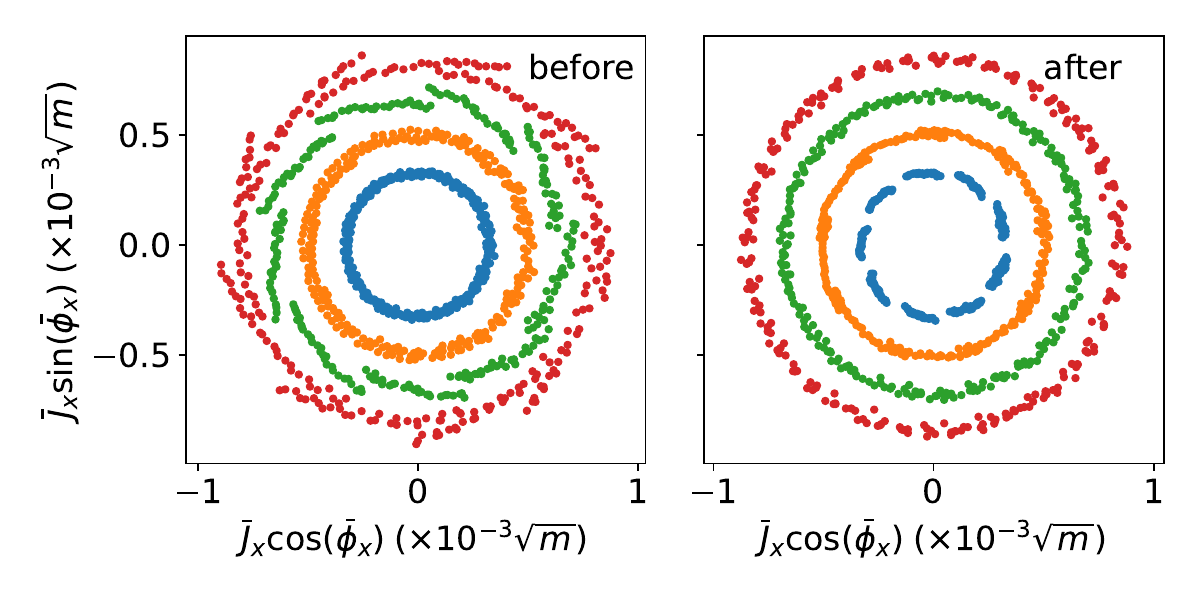}
     \caption{Measured Poincar\'e maps in the $J-\phi$ plane before (in the left) and after (in the right) the optimization. Entropy reduction is the result of a regularization on the phase space trajectories by mitigating the distortion in the vicinity of $\nu_x=\frac{2}{9}$ resonance.}
     \label{fig:PhaseOpt}
 \end{figure}

  The online chaos minimization is accomplished with a Gaussian process optimizer~\cite{williams2006,scikit2011,roussel2024bayesian}. By varying the power supplies of the sextupoles within certain ranges, the optimizer seeks an optimal sextupole configuration, which yield the minimum average entropy over four different amplitudes,
  \begin{equation}
      \overline{S}=\sum_{i=1}^4{\frac{S_{i}}{S_{0,i}}},
  \end{equation}
here $i$ is the index of kick amplitudes, $S_{0}$ represents the entropy of the initial lattice configuration used as the reference. The reference is used here because we would like to see the relative entropy change. Once the entropy converges within its measurement precision, the optimization is terminated. The optimal configuration is checked by re-measuring the Poincar\'e maps and comparing against the initial settings. With the optimal sextupole configuration, the entropy at different amplitudes reduces simultaneously as illustrated in Fig.~\ref{fig:ShenOpt}. A significant improvement on the regularity of the maps is achieved as shown in Fig.~\ref{fig:PhaseOpt}. Particularly the irregular motion in the vicinity of $\nu_x=\frac{2}{9}$ resonance is mitigated. In principle, the reduction of entropy is equivalent to a regularization on the Poincar\'e map. Consequently upon this regularization, the off-axis injection efficiency is improved from 90\% to 100\%.
 
\begin{figure}[!ht]
     \centering
     \includegraphics[width=0.8\columnwidth]{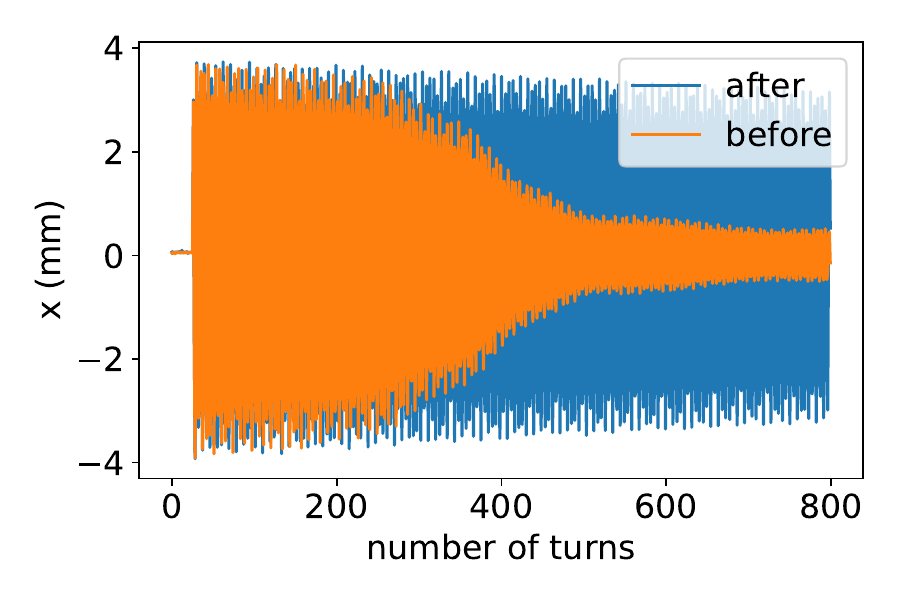}
     \caption{The decoherence of a kicked beam reduces significantly after the optimization. It is the consequence of a smaller TSWA.}
     \label{fig:decoherence}
 \end{figure}

 After the entropy optimization, we also observe a reduction on the TSWA coefficient. The turn-by-turn data as illustrated in Fig.~\ref{fig:decoherence} shows the decoherence is significantly mitigated. Since the radiation damping and linear chromaticity don't change, the fast decay of beam centroid motion before the optimization is mainly due to a large tune-spread generated through the TSWA. With the optimized sextupole configuration, the TSWA becomes smaller, so does the tune-spread inside kicked beams. Then the filamentation process is slowed down. It is another piece of evidence on the improvement of nonlinear dynamics.
 
\section*{Discussions} 
\begin{figure}[!ht]
     \centering
     \includegraphics[width=0.45\columnwidth]{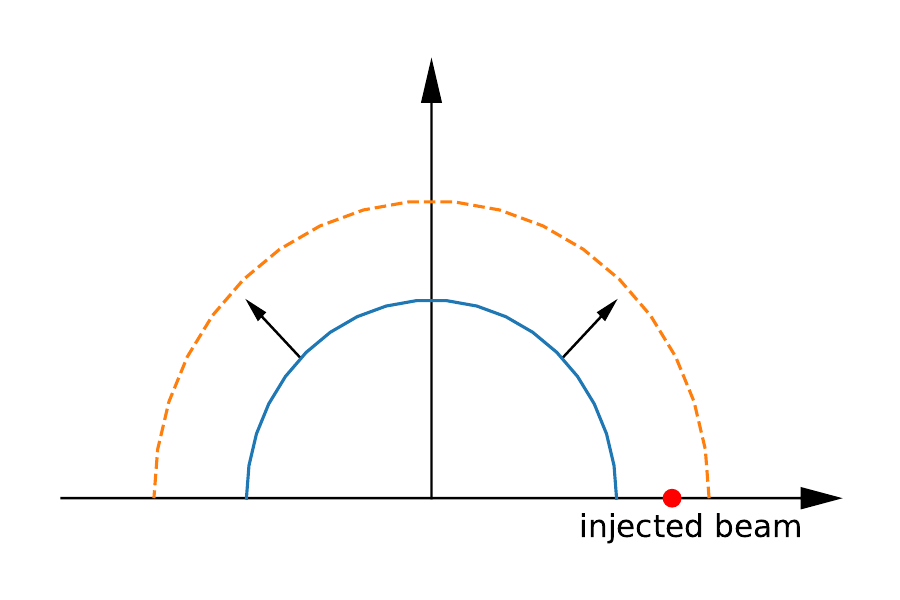}
     \includegraphics[width=0.45\columnwidth]{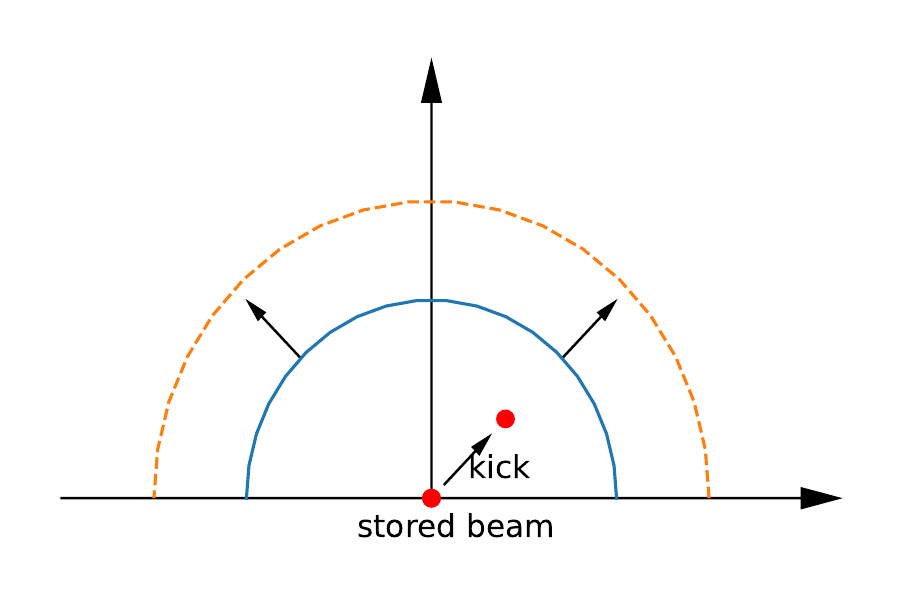}
     \caption{Schematic diagrams of two DA optimization strategies. Left: using injection efficiency of off-axis injected beam as a direct optimization objective. Right: using the entropy of excited stored beams as an indirect optimization objective.}
     \label{fig:rcds_se}
 \end{figure}

 Despite the ultimate target of enlarging the DA, our method differs from another successful method -- injection efficiency maximization~\cite{huang2013algorithm,emery2021application} (IEM) as showing in the schematics in Fig.~\ref{fig:rcds_se}. The blue solid line is an existing DA, and the orange dashed one is more desirable for injection. The enlargement of the DA can be confirmed by the capture efficiency of the injected beam, used as a direct optimization objective in IEM. Our approach uses a measurable chaos indicator with an explicit physics meaning -- the entropy of the excited beam, as the objective. Generally speaking, the particle motion in an accelerator is governed by a nonlinear Hamiltonian $H = H_0+H_1$, where $H_0$ is the linear component, which defines regular motion. $H_1$ is the nonlinearity which can generate chaos if the system is not integrable. By manipulating sextupole configuration in $H_1$, we attempt to let the system have approximate invariants with less chaos around them because it is generally believed that such a $H$ can result in a better DA. Experimentally, our studies confirm that as a result of minimizing the chaos in beam motions, the off-axis injection efficiency is indeed improved in the NSLS-II ring. While various chaos indicators are widely utilized in nonlinear beam dynamics simulations, this is the first time a measurable one is applied to online optimization in a real-world particle accelerator.

 It is worthwhile to mention that the entropy chaos indicator can also be used for design simulation. We test it with the NSLS-II lattice model and obtain some solutions which are comparable to our operation lattice. A detailed analysis of the offline optimization cannot be provided here due to limitations in length.

 Most existing ring-based accelerator lattices (including NSLS-II) are non-integrable, therefore they only have approximate invariants when particle oscillation amplitude is sufficiently small. If a ring lattice is integrable and has analytical invariants~\cite{danilov2010nonlinear}, the near-zero Shannon entropy might be found through some canonical transformations and be potentially used as the experimental criteria of its integrability. Some further exploration on applying Shannon entropy analysis to nonlinear dynamics might be interesting.
 
\section*{Acknowledgments}
 We would like to thank Sergei Nagaitsev and Ivan Morozov for their in-depth discussion. This research is supported by the U.S. Department of Energy under Contract No. DE-SC0012704, the NSLS-II Facility Improvement Projects and the DOE HEP award DE-SC0023722. Coauthor KA is partially sponsored by Brookhaven National Laboratory Graduate Research Internship Program 2024.

\section*{Author contributions}
 YL proposed and developed the method, performed the experimental study and prepared the manuscript. DX and YH developed the method and prepared the manuscript. KA and MS performed the experimental study.  KH and YH supported the experimental study. RR prepared the manuscript. VS and TS sponsored KA's internship.

\section*{Competing interests}
The authors declare no competing interests.

\bibliography{ref.bib}

\begin{thebibliography}{26}%
\makeatletter
\providecommand \@ifxundefined [1]{%
 \@ifx{#1\undefined}
}%
\providecommand \@ifnum [1]{%
 \ifnum #1\expandafter \@firstoftwo
 \else \expandafter \@secondoftwo
 \fi
}%
\providecommand \@ifx [1]{%
 \ifx #1\expandafter \@firstoftwo
 \else \expandafter \@secondoftwo
 \fi
}%
\providecommand \natexlab [1]{#1}%
\providecommand \enquote  [1]{``#1''}%
\providecommand \bibnamefont  [1]{#1}%
\providecommand \bibfnamefont [1]{#1}%
\providecommand \citenamefont [1]{#1}%
\providecommand \href@noop [0]{\@secondoftwo}%
\providecommand \href [0]{\begingroup \@sanitize@url \@href}%
\providecommand \@href[1]{\@@startlink{#1}\@@href}%
\providecommand \@@href[1]{\endgroup#1\@@endlink}%
\providecommand \@sanitize@url [0]{\catcode `\\12\catcode `\$12\catcode
  `\&12\catcode `\#12\catcode `\^12\catcode `\_12\catcode `\%12\relax}%
\providecommand \@@startlink[1]{}%
\providecommand \@@endlink[0]{}%
\providecommand \url  [0]{\begingroup\@sanitize@url \@url }%
\providecommand \@url [1]{\endgroup\@href {#1}{\urlprefix }}%
\providecommand \urlprefix  [0]{URL }%
\providecommand \Eprint [0]{\href }%
\providecommand \doibase [0]{http://dx.doi.org/}%
\providecommand \selectlanguage [0]{\@gobble}%
\providecommand \bibinfo  [0]{\@secondoftwo}%
\providecommand \bibfield  [0]{\@secondoftwo}%
\providecommand \translation [1]{[#1]}%
\providecommand \BibitemOpen [0]{}%
\providecommand \bibitemStop [0]{}%
\providecommand \bibitemNoStop [0]{.\EOS\space}%
\providecommand \EOS [0]{\spacefactor3000\relax}%
\providecommand \BibitemShut  [1]{\csname bibitem#1\endcsname}%
\let\auto@bib@innerbib\@empty
\bibitem [{\citenamefont {Bazzani}\ \emph {et~al.}(2023)\citenamefont
  {Bazzani}, \citenamefont {Giovannozzi}, \citenamefont {Montanari},\ and\
  \citenamefont {Turchetti}}]{bazzani2023}%
  \BibitemOpen
  \bibfield  {author} {\bibinfo {author} {\bibfnamefont {A}~\bibnamefont
  {Bazzani}}, \bibinfo {author} {\bibfnamefont {M}~\bibnamefont {Giovannozzi}},
  \bibinfo {author} {\bibfnamefont {CE}~\bibnamefont {Montanari}}, \ and\
  \bibinfo {author} {\bibfnamefont {G}~\bibnamefont {Turchetti}},\ }\bibfield
  {title} {\enquote {\bibinfo {title} {Performance analysis of indicators of
  chaos for nonlinear dynamical systems},}\ }\href@noop {} {\bibfield
  {journal} {\bibinfo  {journal} {Physical Review E}\ }\textbf {\bibinfo
  {volume} {107}},\ \bibinfo {pages} {064209} (\bibinfo {year}
  {2023})}\BibitemShut {NoStop}%
\bibitem [{\citenamefont {Laskar}(2003)}]{laskar2003frequency}%
  \BibitemOpen
  \bibfield  {author} {\bibinfo {author} {\bibfnamefont {Jacques}\ \bibnamefont
  {Laskar}},\ }\bibfield  {title} {\enquote {\bibinfo {title} {Frequency map
  analysis and particle accelerators},}\ }in\ \href@noop {} {\emph {\bibinfo
  {booktitle} {Proceedings of the 2003 Particle Accelerator Conference}}},\
  Vol.~\bibinfo {volume} {1}\ (\bibinfo {organization} {IEEE},\ \bibinfo {year}
  {2003})\ pp.\ \bibinfo {pages} {378--382}\BibitemShut {NoStop}%
\bibitem [{\citenamefont {Papaphilippou}(2014)}]{papaphilippou2014detecting}%
  \BibitemOpen
  \bibfield  {author} {\bibinfo {author} {\bibfnamefont {Yannis}\ \bibnamefont
  {Papaphilippou}},\ }\bibfield  {title} {\enquote {\bibinfo {title} {Detecting
  chaos in particle accelerators through the frequency map analysis method},}\
  }\href@noop {} {\bibfield  {journal} {\bibinfo  {journal} {Chaos: An
  Interdisciplinary Journal of Nonlinear Science}\ }\textbf {\bibinfo {volume}
  {24}} (\bibinfo {year} {2014})}\BibitemShut {NoStop}%
\bibitem [{\citenamefont {Todesco}\ \emph {et~al.}(1996)\citenamefont
  {Todesco}, \citenamefont {Giovannozzi},\ and\ \citenamefont
  {Scandale}}]{todesco1996}%
  \BibitemOpen
  \bibfield  {author} {\bibinfo {author} {\bibfnamefont {Ezio}\ \bibnamefont
  {Todesco}}, \bibinfo {author} {\bibfnamefont {Massimo}\ \bibnamefont
  {Giovannozzi}}, \ and\ \bibinfo {author} {\bibfnamefont {Walter}\
  \bibnamefont {Scandale}},\ }\bibfield  {title} {\enquote {\bibinfo {title}
  {Fast indicators of long-term stability},}\ }\href@noop {} {\bibfield
  {journal} {\bibinfo  {journal} {Part. Accel.}\ }\textbf {\bibinfo {volume}
  {55}},\ \bibinfo {pages} {27--36} (\bibinfo {year} {1996})}\BibitemShut
  {NoStop}%
\bibitem [{\citenamefont {Guignard}\ and\ \citenamefont
  {Hagel}(1985)}]{guignard1985}%
  \BibitemOpen
  \bibfield  {author} {\bibinfo {author} {\bibfnamefont {Gilbert}\ \bibnamefont
  {Guignard}}\ and\ \bibinfo {author} {\bibfnamefont {J}~\bibnamefont
  {Hagel}},\ }\bibfield  {title} {\enquote {\bibinfo {title} {Sextupole
  correction and dynamic aperture: numerical and analytical tools},}\
  }\href@noop {} {\bibfield  {journal} {\bibinfo  {journal} {Part. Accel.}\
  }\textbf {\bibinfo {volume} {18}},\ \bibinfo {pages} {129--165} (\bibinfo
  {year} {1985})}\BibitemShut {NoStop}%
\bibitem [{\citenamefont {Cai}(2020)}]{cai2020}%
  \BibitemOpen
  \bibfield  {author} {\bibinfo {author} {\bibfnamefont {Yunhai}\ \bibnamefont
  {Cai}},\ }\bibfield  {title} {\enquote {\bibinfo {title} {Parametrization,
  characterization, and optimization of double-bend achromat cell},}\
  }\href@noop {} {\bibfield  {journal} {\bibinfo  {journal} {Physical Review
  Accelerators and Beams}\ }\textbf {\bibinfo {volume} {23}},\ \bibinfo {pages}
  {034002} (\bibinfo {year} {2020})}\BibitemShut {NoStop}%
\bibitem [{\citenamefont {Dragt}(2020)}]{dragt2020}%
  \BibitemOpen
  \bibfield  {author} {\bibinfo {author} {\bibfnamefont {Alex~J.}\ \bibnamefont
  {Dragt}},\ }\href@noop {} {\emph {\bibinfo {title} {Lie Methods for Nonlinear
  Dynamics with Applications to Accelerator Physics}}}\ (\bibinfo  {publisher}
  {unpublished},\ \bibinfo {year} {2020})\BibitemShut {NoStop}%
\bibitem [{\citenamefont {Li}\ \emph {et~al.}(2021)\citenamefont {Li},
  \citenamefont {Hao}, \citenamefont {Hwang} \emph {et~al.}}]{li2021}%
  \BibitemOpen
  \bibfield  {author} {\bibinfo {author} {\bibfnamefont {Yongjun}\ \bibnamefont
  {Li}}, \bibinfo {author} {\bibfnamefont {Yue}\ \bibnamefont {Hao}}, \bibinfo
  {author} {\bibfnamefont {Kilean}\ \bibnamefont {Hwang}},  \emph {et~al.},\
  }\bibfield  {title} {\enquote {\bibinfo {title} {Fast dynamic aperture
  optimization with forward-reversal integration},}\ }\href@noop {} {\bibfield
  {journal} {\bibinfo  {journal} {Nuclear Instruments and Methods in Physics
  Research Section A: Accelerators, Spectrometers, Detectors and Associated
  Equipment}\ }\textbf {\bibinfo {volume} {988}},\ \bibinfo {pages} {164936}
  (\bibinfo {year} {2021})}\BibitemShut {NoStop}%
\bibitem [{\citenamefont {Poincar{\'e}}(1967)}]{poincare1967new}%
  \BibitemOpen
  \bibfield  {author} {\bibinfo {author} {\bibfnamefont {Henri}\ \bibnamefont
  {Poincar{\'e}}},\ }\href@noop {} {\emph {\bibinfo {title} {New methods of
  celestial mechanics}}},\ \bibinfo {number} {450-451}\ (\bibinfo  {publisher}
  {National Aeronautics and Space Administration},\ \bibinfo {year}
  {1967})\BibitemShut {NoStop}%
\bibitem [{\citenamefont {Shannon}(1948)}]{shannon1948}%
  \BibitemOpen
  \bibfield  {author} {\bibinfo {author} {\bibfnamefont {Claude~Elwood}\
  \bibnamefont {Shannon}},\ }\bibfield  {title} {\enquote {\bibinfo {title} {A
  mathematical theory of communication},}\ }\href@noop {} {\bibfield  {journal}
  {\bibinfo  {journal} {The Bell system technical journal}\ }\textbf {\bibinfo
  {volume} {27}},\ \bibinfo {pages} {379--423} (\bibinfo {year}
  {1948})}\BibitemShut {NoStop}%
\bibitem [{\citenamefont {Letellier}(2006)}]{letellier2006}%
  \BibitemOpen
  \bibfield  {author} {\bibinfo {author} {\bibfnamefont {Christophe}\
  \bibnamefont {Letellier}},\ }\bibfield  {title} {\enquote {\bibinfo {title}
  {Estimating the shannon entropy: Recurrence plots versus symbolic
  dynamics},}\ }\href@noop {} {\bibfield  {journal} {\bibinfo  {journal}
  {Physical review letters}\ }\textbf {\bibinfo {volume} {96}},\ \bibinfo
  {pages} {254102} (\bibinfo {year} {2006})}\BibitemShut {NoStop}%
\bibitem [{\citenamefont {Courant}\ and\ \citenamefont
  {Snyder}(1958)}]{courant1958}%
  \BibitemOpen
  \bibfield  {author} {\bibinfo {author} {\bibfnamefont {Ernest~D}\
  \bibnamefont {Courant}}\ and\ \bibinfo {author} {\bibfnamefont {Hartland~S}\
  \bibnamefont {Snyder}},\ }\bibfield  {title} {\enquote {\bibinfo {title}
  {Theory of the alternating-gradient synchrotron},}\ }\href@noop {} {\bibfield
   {journal} {\bibinfo  {journal} {Annals of physics}\ }\textbf {\bibinfo
  {volume} {3}},\ \bibinfo {pages} {1--48} (\bibinfo {year}
  {1958})}\BibitemShut {NoStop}%
\bibitem [{\citenamefont {Edwards}\ and\ \citenamefont
  {Teng}(1973)}]{edwards1973}%
  \BibitemOpen
  \bibfield  {author} {\bibinfo {author} {\bibfnamefont {DA}~\bibnamefont
  {Edwards}}\ and\ \bibinfo {author} {\bibfnamefont {LC}~\bibnamefont {Teng}},\
  }\bibfield  {title} {\enquote {\bibinfo {title} {Parametrization of linear
  coupled motion in periodic systems},}\ }\href@noop {} {\bibfield  {journal}
  {\bibinfo  {journal} {IEEE Transactions on nuclear science}\ }\textbf
  {\bibinfo {volume} {20}},\ \bibinfo {pages} {885--888} (\bibinfo {year}
  {1973})}\BibitemShut {NoStop}%
\bibitem [{\citenamefont {Chao}\ \emph {et~al.}(1988)\citenamefont {Chao},
  \citenamefont {Johnson}, \citenamefont {Peggs}, \citenamefont {Peterson},
  \citenamefont {Saltmarsh}, \citenamefont {Schachinger}, \citenamefont
  {Meller}, \citenamefont {Siemann}, \citenamefont {Talman}, \citenamefont
  {Morton} \emph {et~al.}}]{chao1988}%
  \BibitemOpen
  \bibfield  {author} {\bibinfo {author} {\bibfnamefont {A}~\bibnamefont
  {Chao}}, \bibinfo {author} {\bibfnamefont {D}~\bibnamefont {Johnson}},
  \bibinfo {author} {\bibfnamefont {S}~\bibnamefont {Peggs}}, \bibinfo {author}
  {\bibfnamefont {J}~\bibnamefont {Peterson}}, \bibinfo {author} {\bibfnamefont
  {C}~\bibnamefont {Saltmarsh}}, \bibinfo {author} {\bibfnamefont
  {L}~\bibnamefont {Schachinger}}, \bibinfo {author} {\bibfnamefont
  {R}~\bibnamefont {Meller}}, \bibinfo {author} {\bibfnamefont {R}~\bibnamefont
  {Siemann}}, \bibinfo {author} {\bibfnamefont {R}~\bibnamefont {Talman}},
  \bibinfo {author} {\bibfnamefont {P}~\bibnamefont {Morton}},  \emph
  {et~al.},\ }\bibfield  {title} {\enquote {\bibinfo {title} {Experimental
  investigation of nonlinear dynamics in the fermilab tevatron},}\ }\href@noop
  {} {\bibfield  {journal} {\bibinfo  {journal} {Physical review letters}\
  }\textbf {\bibinfo {volume} {61}},\ \bibinfo {pages} {2752} (\bibinfo {year}
  {1988})}\BibitemShut {NoStop}%
\bibitem [{\citenamefont {Bazzani}\ \emph {et~al.}(1994)\citenamefont
  {Bazzani}, \citenamefont {Servizi}, \citenamefont {Turchetti},\ and\
  \citenamefont {Todesco}}]{bazzani1994normal}%
  \BibitemOpen
  \bibfield  {author} {\bibinfo {author} {\bibfnamefont {Armando}\ \bibnamefont
  {Bazzani}}, \bibinfo {author} {\bibfnamefont {G}~\bibnamefont {Servizi}},
  \bibinfo {author} {\bibfnamefont {G}~\bibnamefont {Turchetti}}, \ and\
  \bibinfo {author} {\bibfnamefont {Ezio}\ \bibnamefont {Todesco}},\
  }\href@noop {} {\emph {\bibinfo {title} {A normal form approach to the theory
  of nonlinear betatronic motion}}},\ \bibinfo {number} {CERN-94-02}\ (\bibinfo
   {publisher} {CERN},\ \bibinfo {year} {1994})\BibitemShut {NoStop}%
\bibitem [{\citenamefont {Meller}\ \emph {et~al.}(1987)\citenamefont {Meller},
  \citenamefont {Chao}, \citenamefont {Peterson}, \citenamefont {Peggs},\ and\
  \citenamefont {Furman}}]{meller1987decoherence}%
  \BibitemOpen
  \bibfield  {author} {\bibinfo {author} {\bibfnamefont {RE}~\bibnamefont
  {Meller}}, \bibinfo {author} {\bibfnamefont {AW}~\bibnamefont {Chao}},
  \bibinfo {author} {\bibfnamefont {JM}~\bibnamefont {Peterson}}, \bibinfo
  {author} {\bibfnamefont {Stephen~G}\ \bibnamefont {Peggs}}, \ and\ \bibinfo
  {author} {\bibfnamefont {M}~\bibnamefont {Furman}},\ }\bibfield  {title}
  {\enquote {\bibinfo {title} {Decoherence of kicked beams},}\ }\href@noop {}
  {\bibfield  {journal} {\bibinfo  {journal} {SSCN-360}\ } (\bibinfo {year}
  {1987})}\BibitemShut {NoStop}%
\bibitem [{\citenamefont {Lee}(1991)}]{lee1991decoherence}%
  \BibitemOpen
  \bibfield  {author} {\bibinfo {author} {\bibfnamefont {SY}~\bibnamefont
  {Lee}},\ }\bibfield  {title} {\enquote {\bibinfo {title} {Decoherence of
  kicked beams ii},}\ }\href@noop {} {\bibfield  {journal} {\bibinfo  {journal}
  {Internal Rep. SSCL-N-749}\ } (\bibinfo {year} {1991})}\BibitemShut {NoStop}%
\bibitem [{\citenamefont {Hao}(1991)}]{hao1991symbolic}%
  \BibitemOpen
  \bibfield  {author} {\bibinfo {author} {\bibfnamefont {Bai-lin}\ \bibnamefont
  {Hao}},\ }\bibfield  {title} {\enquote {\bibinfo {title} {Symbolic dynamics
  and characterization of complexity},}\ }\href@noop {} {\bibfield  {journal}
  {\bibinfo  {journal} {Physica D: Nonlinear Phenomena}\ }\textbf {\bibinfo
  {volume} {51}},\ \bibinfo {pages} {161--176} (\bibinfo {year}
  {1991})}\BibitemShut {NoStop}%
\bibitem [{\citenamefont {Dierker}(2007)}]{dierker2007}%
  \BibitemOpen
  \bibfield  {author} {\bibinfo {author} {\bibfnamefont {Steve}\ \bibnamefont
  {Dierker}},\ }\bibfield  {title} {\enquote {\bibinfo {title} {{NSLS-II}
  preliminary design report},}\ }\href {https://www.osti.gov/biblio/1010602}
  {\bibfield  {journal} {\bibinfo  {journal} {Brookhaven National Laboratory}\
  } (\bibinfo {year} {2007})}\BibitemShut {NoStop}%
\bibitem [{\citenamefont {Li}\ \emph {et~al.}(2024)\citenamefont {Li},
  \citenamefont {Ha}, \citenamefont {Padrazo} \emph {et~al.}}]{li2024}%
  \BibitemOpen
  \bibfield  {author} {\bibinfo {author} {\bibfnamefont {Yongjun}\ \bibnamefont
  {Li}}, \bibinfo {author} {\bibfnamefont {Kiman}\ \bibnamefont {Ha}}, \bibinfo
  {author} {\bibfnamefont {Danny}\ \bibnamefont {Padrazo}},  \emph {et~al.},\
  }\bibfield  {title} {\enquote {\bibinfo {title} {Dedicated beam position
  monitor pair for model-independent lattice characterization at nsls-ii},}\
  }\href@noop {} {\bibfield  {journal} {\bibinfo  {journal} {Nuclear
  Instruments and Methods in Physics Research Section A: Accelerators,
  Spectrometers, Detectors and Associated Equipment}\ }\textbf {\bibinfo
  {volume} {1065}},\ \bibinfo {pages} {169557} (\bibinfo {year}
  {2024})}\BibitemShut {NoStop}%
\bibitem [{\citenamefont {Williams}\ and\ \citenamefont
  {Rasmussen}(2006)}]{williams2006}%
  \BibitemOpen
  \bibfield  {author} {\bibinfo {author} {\bibfnamefont {Christopher~KI}\
  \bibnamefont {Williams}}\ and\ \bibinfo {author} {\bibfnamefont
  {Carl~Edward}\ \bibnamefont {Rasmussen}},\ }\href@noop {} {\emph {\bibinfo
  {title} {Gaussian processes for machine learning}}},\ Vol.~\bibinfo {volume}
  {2}\ (\bibinfo  {publisher} {MIT press Cambridge, MA},\ \bibinfo {year}
  {2006})\BibitemShut {NoStop}%
\bibitem [{\citenamefont {Pedregosa}\ \emph {et~al.}(2011)\citenamefont
  {Pedregosa} \emph {et~al.}}]{scikit2011}%
  \BibitemOpen
  \bibfield  {author} {\bibinfo {author} {\bibfnamefont {F.}~\bibnamefont
  {Pedregosa}} \emph {et~al.},\ }\bibfield  {title} {\enquote {\bibinfo {title}
  {Scikit-learn: Machine learning in {P}ython},}\ }\href@noop {} {\bibfield
  {journal} {\bibinfo  {journal} {Journal of Machine Learning Research}\
  }\textbf {\bibinfo {volume} {12}},\ \bibinfo {pages} {2825--2830} (\bibinfo
  {year} {2011})}\BibitemShut {NoStop}%
\bibitem [{\citenamefont {Roussel}\ \emph {et~al.}(2024)\citenamefont {Roussel}
  \emph {et~al.}}]{roussel2024bayesian}%
  \BibitemOpen
  \bibfield  {author} {\bibinfo {author} {\bibfnamefont {R.}~\bibnamefont
  {Roussel}} \emph {et~al.},\ }\bibfield  {title} {\enquote {\bibinfo {title}
  {Bayesian optimization algorithms for accelerator physics},}\ }\href@noop {}
  {\bibfield  {journal} {\bibinfo  {journal} {Physical Review Accelerators and
  Beams}\ }\textbf {\bibinfo {volume} {27}},\ \bibinfo {pages} {084801}
  (\bibinfo {year} {2024})}\BibitemShut {NoStop}%
\bibitem [{\citenamefont {Huang}\ \emph {et~al.}(2013)\citenamefont {Huang},
  \citenamefont {Corbett}, \citenamefont {Safranek},\ and\ \citenamefont
  {Wu}}]{huang2013algorithm}%
  \BibitemOpen
  \bibfield  {author} {\bibinfo {author} {\bibfnamefont {Xiaobiao}\
  \bibnamefont {Huang}}, \bibinfo {author} {\bibfnamefont {Jeff}\ \bibnamefont
  {Corbett}}, \bibinfo {author} {\bibfnamefont {James}\ \bibnamefont
  {Safranek}}, \ and\ \bibinfo {author} {\bibfnamefont {Juhao}\ \bibnamefont
  {Wu}},\ }\bibfield  {title} {\enquote {\bibinfo {title} {An algorithm for
  online optimization of accelerators},}\ }\href@noop {} {\bibfield  {journal}
  {\bibinfo  {journal} {Nuclear Instruments and Methods in Physics Research
  Section A: Accelerators, Spectrometers, Detectors and Associated Equipment}\
  }\textbf {\bibinfo {volume} {726}},\ \bibinfo {pages} {77--83} (\bibinfo
  {year} {2013})}\BibitemShut {NoStop}%
\bibitem [{\citenamefont {Emery}\ \emph {et~al.}(2021)\citenamefont {Emery},
  \citenamefont {Shang}, \citenamefont {Sun},\ and\ \citenamefont
  {Huang}}]{emery2021application}%
  \BibitemOpen
  \bibfield  {author} {\bibinfo {author} {\bibfnamefont {Louis}\ \bibnamefont
  {Emery}}, \bibinfo {author} {\bibfnamefont {Hairong}\ \bibnamefont {Shang}},
  \bibinfo {author} {\bibfnamefont {Yipeng}\ \bibnamefont {Sun}}, \ and\
  \bibinfo {author} {\bibfnamefont {Xiaobiao}\ \bibnamefont {Huang}},\
  }\bibfield  {title} {\enquote {\bibinfo {title} {Application of a machine
  learning based algorithm to online optimization of the nonlinear beam
  dynamics of the {Argonne Advanced Photon Source}},}\ }\href@noop {}
  {\bibfield  {journal} {\bibinfo  {journal} {Physical Review Accelerators and
  Beams}\ }\textbf {\bibinfo {volume} {24}},\ \bibinfo {pages} {082802}
  (\bibinfo {year} {2021})}\BibitemShut {NoStop}%
\bibitem [{\citenamefont {Danilov}\ and\ \citenamefont
  {Nagaitsev}(2010)}]{danilov2010nonlinear}%
  \BibitemOpen
  \bibfield  {author} {\bibinfo {author} {\bibfnamefont {V}~\bibnamefont
  {Danilov}}\ and\ \bibinfo {author} {\bibfnamefont {S}~\bibnamefont
  {Nagaitsev}},\ }\bibfield  {title} {\enquote {\bibinfo {title} {Nonlinear
  accelerator lattices with one and two analytic invariants},}\ }\href@noop {}
  {\bibfield  {journal} {\bibinfo  {journal} {Physical Review Special
  Topics—Accelerators and Beams}\ }\textbf {\bibinfo {volume} {13}},\
  \bibinfo {pages} {084002} (\bibinfo {year} {2010})}\BibitemShut {NoStop}%
\end{thebibliography}%

\end{document}